\documentclass[24pt]{article}
\usepackage{amssymb}
\usepackage{amsfonts}
\usepackage{amssymb,amsmath}
\usepackage{mathrsfs}
\usepackage{latexsym}
\usepackage{amsmath}
\usepackage{latexsym}
\usepackage[cp1251]{inputenc}
\usepackage{graphicx}
\usepackage{color}

\title{Quantum wave impedance calculation for an arbitrary piesewise constant potential}
\author{O. I. Hryhorchak\\
{\small Department for Theoretical Physics, Ivan Franko National
University of Lviv,}\\
{\small 12, Drahomanov Str., Lviv, UA--79005,
Ukraine}\\
\small{\it{Orest.Hryhorchak@lnu.edu.ua}}}

\def\ch{\mathop{\rm ch}\nolimits}
\def\sh{\mathop{\rm sh}\nolimits}
\def\th{\mathop{\rm th}\nolimits}
\def\cth{\mathop{\rm cth}\nolimits}

\def\const{\mathop{\rm const}\nolimits}

\def\arctanh{\mathop{\rm arctanh}\nolimits}

\begin{document}
\renewcommand{\abstractname}{Abstract}
\maketitle

\begin{abstract}
The method of a determination of a quantum wave impedance for an arbitrary piecewise constant potential was developed. On the base of this method both the well-known iterative formula \cite{Khondker_Khan_Anwar:1988} and alternative ways for a quantum wave impedance calculation were derived. A scattering and a bound state case were considered.  The general form of a wave function of bound states for an arbitrary piesewise constant potential was obtained as well as transmission and  reflection coefficients in a scattering case. The applying of method was demonstarted on the system of double well/barrier. 
\end{abstract}

\section{Introduction}
In the previous papers \cite{Arx1:2020, Arx2:2020} it was demonstrated that using a quantum wave impedance approach 
allows simplifying (in comparison with other approaches) and generalizing  the process of single barrier studying significantly. Papers  \cite{Nelin:2007, Bojko_Berezjanskyi_Nelin:2007, Nelin_Vodolazka:2014, Akhmedov_Nelin:2007} are dedicated to the same topic. 

But this applies not just for single barrier systems and the quantum wave impedance approach can be generalized to the sequence of arbitrary rectangular barriers/wells. It is worth to say that the consideration of such model systems, namely systems of barriers/wells, plays an important role in studying nanostructures. For example in the article \cite{Nelin:2009} on the base of a model for a transmission line the analytical expressions for resonant parameters and characteristics of typical barrier structures of nanoelect\-ronics were obtained while paper \cite{Nelin:2010} deals with the radioengineering and optic models in nanoelectronics and describes the application of radioengineering and optic models to nanoelectronic problems. Analytical expressions for resonant parameters and characteristics of typical barrier structures are presented in that article. In \cite{Akhmedov_Nelin:2007}
the application of the impedance model for
nanoelectronic quantum-mechanical structures modelling is described. There are a lot of other papers  \cite{Nelin:2008,  Nazarko_etall:2010, Nelin:2011, Nelin:2009_1, Nelin_Sergiyenko:2008, Gindikina_etall:2015, Vodolazka_Mikolaychik_Nelin:2017} in which the investigation of  systems of barriers/wells were done by the quantum wave impedance method. 

But usually only systems of two or three barriers/wells are considered analytically \cite{Akhmedov_Nelin:2007, Nelin:2009_1, Gindikina_etall:2015, Vodolazka_Mikolaychik_Nelin:2017} since the more barriers/wells are included into the consideretion the more task is complicated. And the complexity of calculations increases non-lineary. 

The aim of this article is to develop the method for a determination of a quantum wave impedance for an arbitrary piecewise constant potential. It will allow  obtaining both the well-known iterative formula \cite{Khondker_Khan_Anwar:1988} which was transfered from the tramsmission line theory and other more easy ways of a quantum wave impedance calculation.

 \section{General consideration}
 Assume that we have a particle subjected to a piecewise constant potential of the following form:
 \begin{eqnarray}\label{pw_pot}
 U(x)=U_0\theta(x_0-x)+\sum_{i=1}^{N} U_i[\theta(x-x_{i-1})-\theta(x-x_{i})]+
U_{N+1}\theta(x-x_{N+1}),
 \end{eqnarray}
 where $\theta(x)$ is a Heaviside step function, $U_i=\const$, $i=0\ldots N\!+\!1$, $x_0<x_1<\ldots<x_{N+1}$. On the base of the solution \cite{Arx2:2020} for a constant potential we assume that the solution of the equation for a quantum wave impedance function \cite{Arx1:2020} with a potential (\ref{pw_pot}) has the following form:
 \begin{eqnarray}\label{Z_pw_as}
 Z(x)&=&z_0\th\left[\gamma_0x+\phi_0\right]\theta(x_0-x)+\sum_{i=1}^{N} z_i
 \th\left[\gamma_ix+\phi_i\right][\theta(x-x_{i-1})-\theta(x-x_{i})]+\nonumber\\
 &+&z_{N+1}\th\left[\gamma_{N+1}x+\phi_{N+1}\right]\theta(x-x_{N+1}).
 \end{eqnarray}
 Substitution of (\ref{Z_pw_as}) into  the equation for a quantum wave impedance function \cite{Arx1:2020} gives
 \begin{eqnarray}
 &&\frac{z_0\gamma_0\theta(x_0-x)}{\ch^2\left[\gamma_0x+\phi_0\right]}-z_0
 \th\left[\gamma_0x+\phi_0\right]\delta(x_0-x)+\nonumber\\
 &+&\sum_{i=1}^{N} \frac{z_i\gamma_i[\theta(x-x_{i-1})-\theta(x-x_{i})]}{\ch^2\left(\gamma_ix+\phi_i\right)}+z_i\th\left[\gamma_ix+\phi_i\right]\times\nonumber\\
 &\times& [\delta(x-x_{i-1})-\delta(x-x_{i})]
 + \frac{z_{N+1}\gamma_{N+1}\theta(x-x_{N+1})}{\ch^2\left[\gamma_{N+1}x+\phi_{N+1})\right]}+\nonumber\\
 &+&z_{N+1}\th\left[\gamma_{N+1}x+\phi_{N+1}\right]
 \delta(x-x_{N+1})+\nonumber\\
 &+&z_0^2\th^2\left[\gamma_0x+\phi_0\right]\theta(x_0-x)
 +i\frac{m}{\hbar}\sum_{i=1}^{N} z_i^2\th^2\left[\gamma_ix+\phi_i\right]\times\nonumber\\
 &\times&[\theta(x-x_{i-1})-\theta(x-x_{i})]
 +z_{N+1}^2\th^2\left[\gamma_{N+1}x+\phi_{N+1}\right]\times\nonumber\\
 &\times&\theta(x-x_{N+1})
 =i\frac{2}{\hbar}\left(E-U_0\theta(x_0-x)\frac{}{}\right.-\nonumber\\
 &-&\left.\sum_{i=1}^{N} U_i[\theta(x-x_{i-1})
 -\theta(x-x_{i})]
 +U_{N+1} \theta(x-x_{N+1})\right)\!\!.
 \end{eqnarray}
 To provide this equation to be valid we have to put
 \begin{eqnarray}\label{kizi}
 \gamma_i=i\frac{m}{\hbar}z_i,\quad
 i\frac{m}{\hbar}z_i^2=i\frac{2}{\hbar}\left(E-U_i\right),
 \end{eqnarray}
 \begin{eqnarray}\label{iter}
 z_i\th\left[\gamma_ix_i+\phi_i\right]=z_{i+1}\th\left[\gamma_{i+1}x_i+\phi_{i+1}\right].
 \end{eqnarray}
 The system of equations (\ref{iter}) is the expression of a continuity of a quantum wave impedance function and it relates to the iterative process of a quantum wave impedance calculation. To demonstrate it let's write two arbitrary consecutive equations of the system (\ref{iter}):
 \begin{eqnarray}
 &&-z_{m-1}\th\left[\gamma_{m-1}x_{m-1}+\phi_{m-1}\right]+z_m\th\left[\gamma_mx_{m-1}+\phi_m\right]=0,\nonumber\\
 &&-z_{m}\th\left[\gamma_{m}x_{m}+\phi_{m}\right]+z_{m+1}\th\left[\gamma_{m+1}x_{m}+\phi_{m+1}\right]=0.
 \end{eqnarray}
 Our task is to find a relation between $Z(x_{m-1})$ and $Z(x_{m})$, where
 \begin{eqnarray} Z(x_m)&=&z_{m}\th\left[\gamma_{m}x_{m}+\phi_{m}\right],\nonumber\\
 Z(x_{m-1})&=&z_{m-1}\th\left[\gamma_{m-1}x_{m-1}+\phi_{m-1}\right].
 \end{eqnarray}
 For this we express $\th\left[\gamma_{m}x_{m}+\phi_{m}\right]$ from the second equation and substitute it into the first one. After simple transformations we get:
 \begin{eqnarray}
 Z(x_{m-1})=
 z_m\frac{Z(x_{m})-z_m\th\left[\gamma_{m}\Delta x_m\right]}{z_m-Z(x_{m})\th\left[\gamma_{m}\Delta x_m\right]},
 \end{eqnarray}
 where 
 \begin{eqnarray}
 \Delta x_m=x_m-x_{m-1}.
 \end{eqnarray}
 It is in fact the well-known formula \cite{Khondker_Khan_Anwar:1988}.
 
 In a case of a wave propagating in the opposite direction we have the similar expression
 \begin{eqnarray}\label{Z_iter_r}
 Z(x_{m})=
 z_m\frac{Z(x_{m-1})+z_m\th\left[\gamma_{m}\Delta x_m\right]}{z_m+Z(x_{m-1})\th\left[\gamma_{m}\Delta x_m\right]}.
 \end{eqnarray}

 \section{Scattering case for a piecewise constant po\-tential}
 
 The relation (\ref{iter}), which we got in the previous section is in fact the system of $N+1$ equations for finding $N+2$ values of $\phi_i$, $i = 0\ldots{N+1}$ as functions of energy $E$. So, having $N+1$ equations we have $N+3$ undefined values: energy $E$ and $N+2$ of $\phi_i$. Thus, to solve the system of equations (\ref{iter}) 
 unambiguously we have to join one (scattering case) or two equations (bound states case or case of resonant levels) which relate mentioned values based on the properties of a wave-function. 
 It is an expected situation since besides of a differential equation we need to have boundary conditions for getting an unambiguous solution.
 
 On the base of results obtained in \cite{Arx1:2020} in the  scattering case we have to put $\phi_0=\infty$ for a wave incidenting on the left and $\phi_{N+1}=-\infty$ for a wave which moves in the opposite direction. In this case the value of $\exp[-2\phi_{N+1}]$ (or $\exp[2\phi_0]$) will determine a wave reflection amplitude coefficient as a function of an energy. By setting $r_{N+1}=0$ we find resonant levels. Generally by setting one of the parameters $r_0$ or  $r_{N+1}$ equal to zero we define the direction of a wave propagation and by setting both $r_0$ and $r_{N+1}$ equal to zero we get the condition for finding energies of resonant levels. 
 
 Using (\ref{iter}) we can also find an expression for $\th[\phi_m]$ and for a wave reflection amplitude coefficient $r_m$ at points $x_m$ :
 \begin{eqnarray}
 \th[\phi_m]=\frac{z_m\th(\gamma_mx_m)-Z(x_m)}{z_m-Z(x_m)\th(\gamma_mx_m)},
 \end{eqnarray}
 \begin{eqnarray}
 r_m=\exp[-2\phi_m]=\exp[2\gamma_mx_m]\frac{z_m-Z(x_m)}{z_m+Z(x_m)}.
 \end{eqnarray}
 Notice that here $r_m=\frac{B_m}{A_m}$, where $A_m$ is an amplitude of a wave which propagates to the right in the region $x\in (x_{m-1}..x_m)$ and $B_m$ is the amplitude of a reflected at point $x=x_m$ wave (which before reflection had an amplitude $A_m$).

 \section{Bound states case for a piecewise constant po\-tential} 
 
 In the bound states case ($E<U_0$, $E<U_{N+1}$) we have to put both  $\phi_0=\infty$ and  
 $\phi_{N+1}=-\infty$ in the (\ref{iter}) system. It gives us the possibility to find both energies of eigenstates and values of phases $\phi_i$, $i=1\ldots N$:
 \begin{eqnarray}\label{eigen_sys}
 &&-z_0=z_1\th\left[\gamma_1x_0+\phi_1\right],\nonumber\\
 &&z_i\th\left[\gamma_ix_i+\phi_i\right]=z_{i+1}\th\left[\gamma_{i+1}x_i+\phi_{i+1}\right], i=1\ldots N,\nonumber\\
 &&z_{N}\th\left[\gamma_{N}x_{N}+\phi_{N}\right]=z_{N+1}.
 \end{eqnarray}
 Thereby, a quantum wave impedance is fully determined (for bound states):
 \begin{eqnarray}
 Z(x)\!\!\!&=&\!\!\!-z_0\theta(x_0\!-\!x)\!+\!\sum_{i=1}^{N} z_i
 \th\left[\gamma_ix+\phi_i\right][\theta(x-x_{i-1})\!-\!\theta(x-x_{i})]+\nonumber\\
 &+&z_{N+1}\theta(x-x_{N+1}),
 \end{eqnarray}
 where $E$ and $\phi_i$ are the solutions of the system (\ref{eigen_sys}).
 
 Taking into account that 
 \begin{eqnarray}
 \int f(x)\theta(x-a)dx=(F(x)-F(a))\theta(x-a),
 \end{eqnarray}
 where
 \begin{eqnarray}
 F(x)=\int f(x)dx
 \end{eqnarray}
 and using formula which relates a quantum wave impedance function and a wave function \cite{Arx1:2020} we can find wave-functions of bound states in the following form
 \begin{eqnarray}
 \psi(x)\!\!\!&=&\!\!\!A\exp\left(\frac{}{}(x-x_0)\theta(x_0-x)+x_0+\sum_{i=0}^N \ln\frac{\ch\left[\gamma_ix+\phi_i\right]}{\ch\left[\gamma_ix_{i-1}+\phi_i\right]}\theta(x-x_{i-1})-\right.\nonumber\\
 \!\!\!&-&\!\!\!\left.\sum_{i=0}^N\ln\frac{\ch\left[\gamma_ix+\phi_i\right]}{\ch\left[\gamma_ix_i+\phi_i\right]}\theta(x-x_i)
 - (x-x_{N+1})\theta(x-x_{N+1})\frac{}{} \right).
 \end{eqnarray}

 \section{System of two symmetric barriers. An iterative approach}

 In this section we consider a symmetric rectangular double barrier system, the potential energy of which has the following form:
 \begin{eqnarray}\label{Pot_dbs}
 U(x)=U_b\left[\theta(x+b+a)-\theta(x+a)\right]+
U_b\left[\theta(x-a)-\theta(x-b-a)\right],
 \end{eqnarray}
 where $U_b>0$ is a constant. Then using an iterative process with a load impedance $z_a$ at a point $x=a+b$ we, step by step, can find a value for a quantum wave impedance at point $x=-a-b$. So
 \begin{eqnarray}
 Z(a)&=&z_b\frac{z_a\ch[\gamma_bb]-z_b\sh[\gamma_bb]}
 {z_b\ch[\gamma_bb]-z_a\sh[\gamma_bb]},\nonumber\\
 Z(-a)&=&z_a\frac{Z(a+b)\ch[2\gamma_aa]-z_a\sh[2\gamma_aa]}
 {z_0\ch[2\gamma_aa]-Z(a+b)\sh[2\gamma_aa]},\nonumber\\
 Z(-a-b)&=&z_b\frac{Z(-a)\ch[\gamma_bb]-z_b\sh[\gamma_bb]}
 {z_b\ch[\gamma_bb]-Z(-a)\sh[\gamma_bb]},\nonumber\\
 \end{eqnarray}
 where $z_a=\sqrt{2E/m}$, $z_b=\sqrt{2(E-U_b)/m}$. After easy but quite long calculations we get
 \begin{eqnarray}\label{Zdbs}
 Z(-a-b)=z_b\frac{z_a^2z_bG_1+z_b^3G_2-z_az_b^2G_3+z_a^3G_4}{z_b^2z_aG_1+z_a^3G_2-z_bz_a^2G_3+z_b^3G_4},
 \end{eqnarray}
 where
 \begin{eqnarray}
 G_1&=&\ch[2\gamma_bb]\ch[2\gamma_aa]-\ch^2[\gamma_bb]\sh[2\gamma_aa],\nonumber\\ 
 G_2&=&\sh[2\gamma_aa]-\ch^2[\gamma_bb]\sh[2\gamma_aa],\nonumber\\
 G_3&=&\ch^2[\gamma_bb]\sh[2\gamma_aa]-\frac{1}{2}\sh[2\gamma_bb]\sh[2\gamma_aa],\nonumber\\
 G_4&=&\frac{1}{2}\sh[2\gamma_bb]\sh[2\gamma_aa].
 \end{eqnarray}
 Thus, we can find the expression for a transmission probability $T$ in this system:
 \begin{eqnarray}
 &&\!\!\!\!\!\!\!\!T=1-\Biggm|\frac{Z(-a-b)+z_a}{Z(-a-b)-z_a}\Biggm|^2=\nonumber\\
 &&\!\!\!\!\!\!\!\!=1-\Biggm|\!\!\frac{2z_a^2z_b^2G_1+(z_a^4+z_b^4)G_2-z_az_b(z_a^2+z_b^2)(G_3-G_4)}{(z_b^4-z_a^4)G_2-z_az_b(z_b^2-z_a^2)(G_3+G_4)}\!\!\Biggm|^2\!\!\!\!\!.
 \end{eqnarray} 
 Notice that in our case of a double-barrier system $z_a$ and $k_b$ are real and $z_b$ and $k_a$ are imaginary.  
 
 \begin{figure}[h!]
 	\centerline{
 		\includegraphics[clip,scale=1.15]{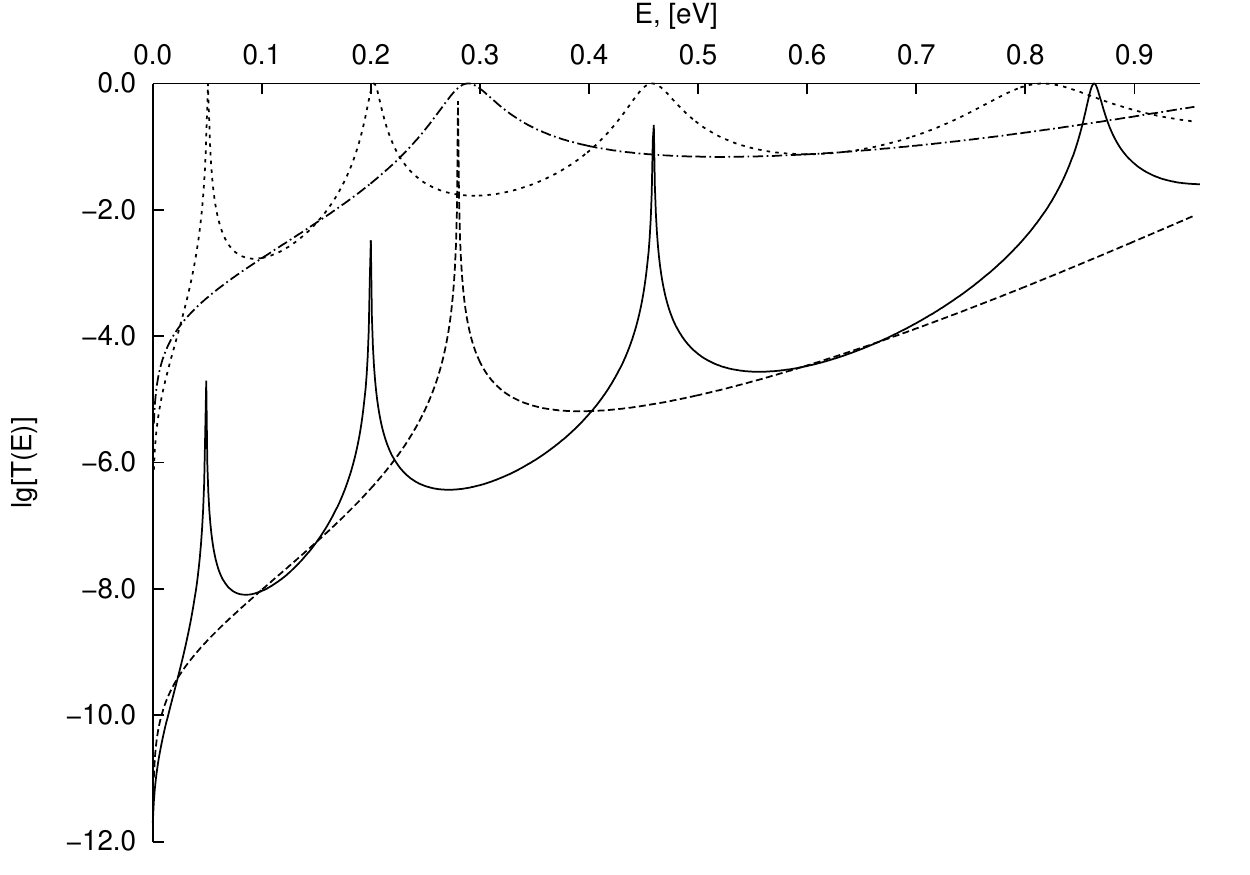}}
 	\caption{\small{Dependence of a transmission probability $T$ on an energy $E$ of a particle for a system of two symmetric rectangular barriers. The height of each barrier is $0.956$ eV \cite{Khondker_Khan_Anwar:1988}. Thickness of each barrier $b$ and distance between them $d=2a$ are represented by different lines. Solid line is for $b=3$ nm, $d=10$ nm; dashed line is for $b=3$ nm, $d=5$ nm; dotted line is for $b=1$ nm, $d=10$ nm and dash-dotted line is for $b=1$ nm, $d=5$ nm. An effective mass of a particle is $m^*=0.1m_0$, where $m_0$ is a ``bare'' mass of an electron.}}
 	\label{fig:U1U2U3}
 \end{figure}
 
 Dependence of a transmission probability $T$ on an energy $E$ of a particle for a system of two symmetric rectangular barriers is represented on Figure 1.
 
 \newpage
 \section{System of two symmetric wells. Quantum wave impe\-dance function approach}
 In a case of a double well system we have the same potential (\ref{Pot_dbs}) but with $U_b<0$ and an obtained in the previous section formula (\ref{Zdbs}) is applicable in this case as well. The condition for eigenstates is as follows
 \begin{eqnarray}
 -z_a=z_b\frac{z_a^2z_bG_1+z_b^3G_2-z_az_b^2G_3+z_a^3G_4}{z_b^2z_aG_1+z_a^3G_2-z_bz_a^2G_3+z_b^3G_4}.
 \end{eqnarray}
 or
 \begin{eqnarray}
 z_a^2z_b^2G_1+(z_a^4+z_b^4)G_2+z_az_b(z_a^2+z_b^2)(G_3+G_4)=0.
 \end{eqnarray}
 Using expressions for $G_1$, $G_2$, $G_3$, $G_4$ after quite exhausting transformations we get
 \begin{eqnarray}
 \frac{(z_a^2-z_b^2)^2\th^2[\gamma_bb]}{z_a^2z_b^2(\th^2[\gamma_bb]\!+\!1)\!-\!z_az_b(z_a^2
 	\!+\!z_b^2)\th[\gamma_bb]} \!=\!\frac{(\th[\gamma_aa]\!-\!1)^2}{\th[\gamma_aa]}.
 \end{eqnarray} 
 Solving this quadratic equation we get two relations for eigenenergies determination:
 \begin{eqnarray}\label{Sol_dws}
 z_a\th[\gamma_aa]&=&z_b\frac{z_a-z_b\th[\gamma_bb]}
 {z_b-z_a\th[\gamma_bb]},\nonumber\\
 z_a\cth[\gamma_aa]&=&z_b\frac{z_a-z_b\th[\gamma_bb]}
 {z_b-z_a\th[\gamma_bb]}.
 \end{eqnarray}
 
 But this way is not optimal and quite complicated. We will show how an application of a quantum wave impedance function can significantly simplify the calculating process . We start by writing matching conditions for a quantum wave impedance function at each of interfaces (it means at points $x=-a-b$, $x=-a$, $x=a$, $x=a+b$):
 \begin{eqnarray}
 &&z_b\th[\gamma_b(-b-a)+\phi_1]=-z_a,\nonumber\\
 &&z_b\th[\gamma_b(-a)+\phi_1]=z_a\th[\gamma_a(-a)+\phi_2],\nonumber\\
 &&z_a\th[\gamma_aa+\phi_2]=z_b\th[\gamma_ba+\phi_3],\nonumber\\
 &&z_b\th[\gamma_b(a+b)+\phi_3]=z_a.
 \end{eqnarray}
 Analysing these relations we see that 
 \begin{eqnarray}
 \phi_1=-\phi_3=\gamma_b(a+b)-\arctanh\left(\frac{z_a}{z_b}\right) 
 \end{eqnarray}
 and that $\phi_2$ is equal to $0+\pi n$ or to $i\frac{\pi}{2}+\pi n$, $n=1,2,3,\ldots$. It gives 
 \begin{eqnarray}
 &&z_b\th[-\gamma_b(b+a)+\phi_1]=-z_a\th[\gamma_aa],\nonumber\\
 &&z_b\th[-\gamma_b(b+a)+\phi_1]=-z_a\cth[\gamma_aa].
 \end{eqnarray} 
 Substituting $\phi_1$ into the previous equations we get
 \begin{eqnarray}
 &&z_b\th\left[\gamma_bb-\arctanh\left(\frac{z_a}{z_b}\right) \right]=-z_a\th[\gamma_aa],\nonumber\\
 &&z_b\th\left[\gamma_bb-\arctanh\left(\frac{z_a}{z_b}\right) \right]=-z_a\cth[\gamma_aa]
 \end{eqnarray}
 or after simple transformations we get the obtained earlier expressions (\ref{Sol_dws}).

\section{Conclusions}
Using a quantum wave impedance concept we have formalized the process of finding eigenenergies and eigenfunctions of bound states for an arbitrary piesewise constant potential. Within our approach we obtained both the well-known formula (which was transfered from an electrical transmission line theory) for an iterative determination of a quantum wave impedance and an alternative ways of its calculation which are based on the relations we obtained in this paper. These results  can be especially valuable for numerical studying quantum mechanical systems of barriers/wells. 

We demonstarted the applying of obtained results to the system of symmetric double barrier/well. It is worth to mention that systems of symmetric double barrier/well are well-studied. There are many papers dedicated to this topic \cite{Guo_etall:1988, Nelin:2009_1, Gindikina_etall:2015, Gindikina_Vodolazka_Nelin:2015, Vodolazka_Mikolaychik_Nelin:2017}. One of the reasons of this fact is that these models are widely used in studying different physical systems since they are both simply enough and ``catch'' the significant effects. For example a symmetric double barrier structure is a basic model for a design of nanoelectronic devises in which the resonant tunnelling effects are significant \cite{Guo_etall:1988}.
At the same time double well structures are used for modelling of molecules, q-bites for a quantum computation, quantum chaos, superconducting quantum interference device (SQUID) \cite{Yang_Chu_Han:2004, Nelin:2010, Dur_Heusler:2014, Ashhab_Nori:2010, Klenov_etall:2011, Bonfim:1998, Silvestrini_Stodolsky:2001, Ruggiero_etall:2006} etc. 

We hope that the developed in this paper aproach allow investigating many barriers/wells systems more easily and effectively.

\renewcommand\baselinestretch{1.0}\selectfont
%\renewcommand{\bibname}{Bibliography} 

%\fancyhead[RE,LO]{\sl Bibliography}

\def\name{\vspace*{-0cm}\LARGE 
	%СПИСОК ВИКОРИСТАНИХ ДЖЕРЕЛ
	Bibliography\thispagestyle{empty}}
\addcontentsline{toc}{chapter}{Bibliography}

{\small

	\bibliographystyle{gost780u}
	%\bibliography{\figsfolder full,add}
	\bibliography{full.bib}
	% insert this in the bbl after	 \begin{thebibliography}{}: \interlinepenalty=10000
	
}

\newpage

\end{document}